\documentclass[aps,prl,twocolumn,superscriptaddress,showpacs]{revtex4-1}
\usepackage{amsmath,amsfonts,amssymb,bm,graphicx,hyperref,color,units}

\newcommand{\revision}[1]{{#1}}

\begin{document}

\title{Bistable Photon Emission from a Solid-State Single-Atom Laser}

\author{Neill Lambert}
\affiliation{CEMS, RIKEN, Saitama, 351-0198, Japan}
\author{Franco Nori}
\affiliation{CEMS, RIKEN, Saitama, 351-0198, Japan}
\affiliation{Department of Physics, University of Michigan, Ann Arbor, MI 48109-1040, USA}
\author{Christian Flindt}
\affiliation{Department of Applied Physics, Aalto University, 00076 Aalto, Finland}

\date{\today}

\begin{abstract}
We predict a bistability in the photon emission from a solid-state single-atom laser comprising a microwave cavity coupled to a voltage-biased double quantum dot. To demonstrate that the single-atom laser is bistable, we evaluate the photon emission statistics and show that the distribution takes the shape of a tilted ellipse. The switching rates of the bistability can be extracted from the electrical current and the shot noise in the quantum dots. This provides a means to control the photon emission statistics by modulating the electronic transport in the quantum dots. Our prediction is robust against moderate electronic decoherence and dephasing and is important for current efforts to realize single-atom lasers with gate-defined quantum dots as the gain medium.
\end{abstract}


\maketitle

\emph{Introduction.---} Double quantum dots (DQD) coupled to electronic reservoirs represent a unique type of quantum technology for explorations of coherent phenomena under non-equilibrium conditions \cite{Wiel2002,Buluta2011,Xiang2013}. Recent breakthrough experiments have demonstrated that quantum dots can be efficiently coupled to microwave resonators forming hybrid light-matter interfaces \cite{Delbecq2011,Delbecq2013,Frey2012,Frey2012b,Petersson2012,Toida2013,Wallraff2013,Viennot2014,Liu2014,Liu2015,Stockklauser2015,Viennot2015,Deng2014}. These advances constitute on-chip realizations of cavity quantum electrodynamics (QED) with electronic conductors \cite{Childress2004,Jin2011,Jin2012,Hu2012,Bergenfeldt2012,Bergenfeldt2013,Lambert2013,Contreras2013,Xu2013a,Xu2013b,Bergenfeldt2014,Kulkarni2014,Cottet2014,Berg2014,Gullans2015}. With the ability to precisely engineer and control the coupling of electronic and photonic degrees of freedom, several applications are now being anticipated. These include the coupling of electronic qubits over macroscopic distances \cite{Childress2004,Bergenfeldt2012,Bergenfeldt2013}, the use of microwave cavities as quantum channels for heat in nano-scale engines \cite{Bergenfeldt2014}, and solid-state implementations of single-atom lasers~\cite{Savage1992,Wang1996}.

Single-atom lasers differ from conventional lasers by operating with only a single emitter and a low number of photons. Such devices harbor a range of non-classical effects, including threshold-free lasing, reversion under large driving, and sub-poisonnian photon statistics~\cite{McKeever2003}. Having been implemented with single atoms in optical cavities, the attention is now shifting towards their realization in solid-state architectures~\cite{Oleg,Ashhab2009}. The role of the single atom is then played by a DQD which interacts with the microwave field of a waveguide resonator. A rich and intricate interplay between the electronic transport in the DQD and the photon emission from the resonator is anticipated, but is not yet understood. A proper understanding of the non-trivial photon emission statistics is key to the operation of solid-state single-atom lasers.

\begin{figure}
\includegraphics[width=\columnwidth]{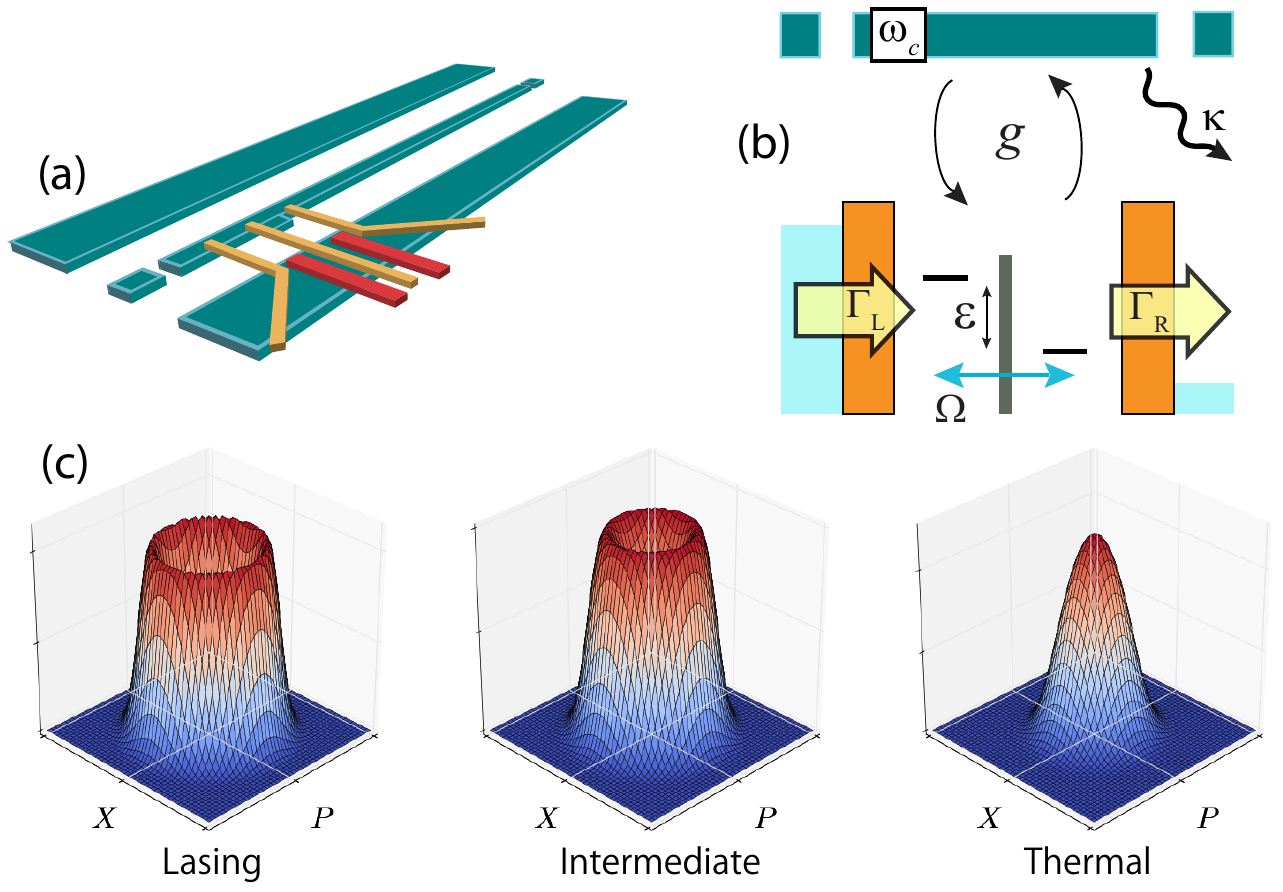}
\caption{(color online). \revision{Solid-state single-atom laser.} (a)~Schematic of a DQD coupled to a microwave cavity forming a hybrid circuit-QED structure. (b) The coupling between the cavity (with frequency $\omega_c$ and decay rate $\kappa$) and the DQD (with tunnel coupling $\Omega$ and dealignment $\varepsilon$) is denoted as~$g$. Single electrons tunnel in and out of the DQD at rates $\Gamma_L$ and $\Gamma_R$, respectively.  The electronic dephasing rate is $\gamma$. (c)~Wigner functions for the cavity photons in the lasing ($\Gamma_R=0.55\omega_c/2\pi$), intermediate ($\Gamma_R=0.84\omega_c/2\pi$), and thermal states ($\Gamma_R=1.25\omega_c/2\pi$). The other parameters are $\varepsilon=\sqrt{(\hbar\omega_c)^2-(2\Omega)^2}$,  $\Omega=0.1\hbar\omega_c$, $\Gamma_L=10\omega_c$, $g=0.05\hbar\omega_c$, $\kappa=0.013\omega_c/2\pi$, and $\gamma=0.05\omega_c/2\pi$.}
\label{fig:1}
\end{figure}

In this Letter we investigate a circuit-QED realization of a single-atom laser for which we predict a bistability in the photon emission statistics. The bistability is not apparent from static observables like the photonic Wigner function. However, as we show, this dynamical feature can be captured by measuring the time-integrated fluctuations of the electrical current and the photon emission statistics. Specifically, \revision{to demonstrate that the single-atom laser is bistable we show that the distribution of emitted photons has the shape of a tilted ellipse.} The switching rates of the bistability can be extracted from the electrical current and shot noise in the DQD which in turn can be used to acquire an exquisite control over the photon emission statistics  \revision{by modulating the electronic transport in the DQD.} The prediction is robust against moderate electronic decoherence and dephasing and may be tested using current technology~\cite{Delbecq2011,Delbecq2013,Frey2012,Frey2012b,Petersson2012,Toida2013,Wallraff2013,Viennot2014,Liu2014,Liu2015,Stockklauser2015,Viennot2015,Deng2014}.

\begin{figure}
\includegraphics[width=0.9\columnwidth]{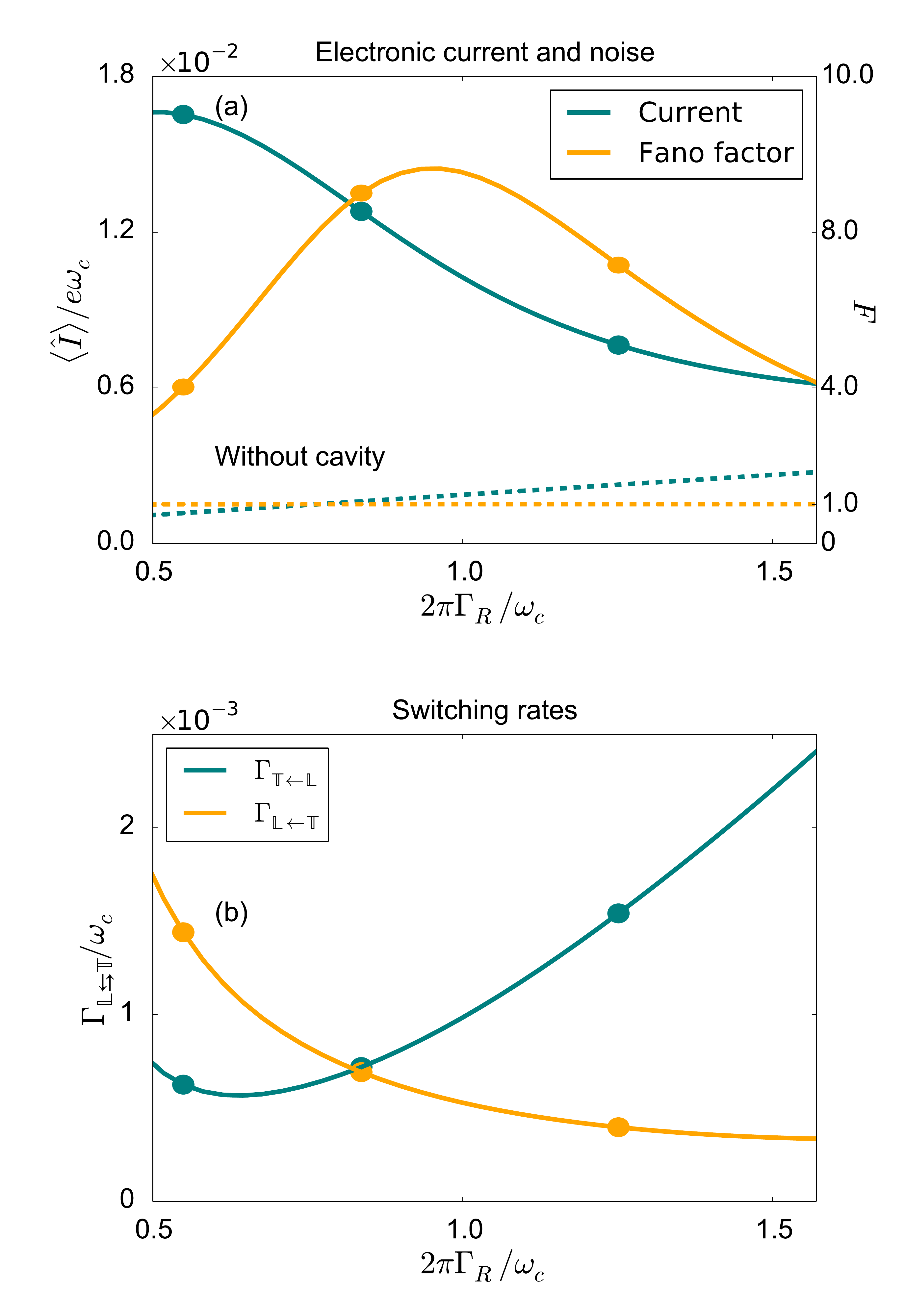}
\caption{(color online). Electrical current, noise, and switching rates. (a) Average electrical current $\langle I\rangle$ and the Fano factor $F=S/e\langle I\rangle$ as functions of the coupling to the right lead. The other parameters are chosen as in~Fig.~\ref{fig:1}. Results without the cavity are shown with dashed lines~\cite{Gurvitz1996}. (b)~Switching rates of the bistability extracted from the current and noise using Eq.~(\ref{eq:switchingrates}). The circles indicate the values of $\Gamma_R$ for which the cavity Wigner functions are shown in Fig.~\ref{fig:1}c and the photon emission statistics are shown in Fig.~\ref{fig:3}.}
\label{fig:2}
\end{figure}

\revision{\emph{Solid-state single-atom laser.---}} Figure~\ref{fig:1} depicts a DQD which is dipole coupled to the microwave field of a waveguide resonator. A large voltage between two electrodes drives an electrical current through the DQD which, due to strong Coulomb interactions, can be occupied by only a single excess electron at the time. The DQD-resonator system is described by the Hamiltonian
\begin{equation}
\hat{H}=\varepsilon\hat{\sigma}_{z}/2+\Omega\hat{\sigma}_{x}+g\hat{\sigma}_{z}\left(\hat{a}^\dag+\hat{a}\right)+\hbar\omega_c\left(\hat{a}^\dag\hat{a}+1/2\right),
\nonumber
\end{equation}
where $\hat{\sigma}_{z}=|L\rangle\!\langle L|-|R\rangle\!\langle R|$ and $\hat{\sigma}_{x}=|L\rangle\!\langle R|+|R\rangle\!\langle L|$ are pseudo-spin operators for the left and right quantum dot levels and $\hat{a}^\dag$ creates photons in the cavity. The first two terms in the Hamiltonian correspond to the dealignment $\varepsilon$ and the tunnel coupling $\Omega$ of the electronic levels. The remaining terms represent the interaction of strength $g$ between an excess electron in the DQD and the electromagnetic field in the cavity with frequency $\omega_c$. We focus below on the resonance $\varepsilon\approx\hbar\omega_c$, where the electronic transport in the DQD pumps photons into the cavity in a manner analogous to a single-atom laser~\cite{Savage1992,McKeever2003}.

To account for the electronic transport and dephasing in the DQD, as well as photonic losses from the cavity, we consider a generalized master equation (GME) for the DQD-resonator density matrix $\hat{\rho}(t)$. With a large bias voltage across the DQD, the GME takes the form~\cite{Gurvitz1996,Lambert2013}
\begin{equation}
\dot{\hat{\rho}}=\mathcal{L}\hat{\rho}=-\frac{i}{\hbar}[\hat{H},\hat{\rho}]+\mathcal{L}_{\mathrm{elec}}\hat{\rho}+\mathcal{L}_{\mathrm{cav}}\hat{\rho},
\label{eq:GME}
\end{equation}
with the commutator corresponding to the coherent evolution of the coupled DQD-resonator system. The incoherent dynamics of the DQD is governed by the term
\begin{equation}
\mathcal{L}_{\mathrm{elec}}\hat{\rho}=\Gamma_{L}\mathcal{D}[\hat{\sigma}_{L}]\hat{\rho}+\Gamma_{R}\mathcal{D}[\hat{\sigma}_{R}^{\dag}]\hat{\rho}
+\gamma\mathcal{D}[\hat{\sigma}_{z}]\hat{\rho}
\end{equation}
which sums up processes where single electrons enter the left quantum dot from the left electrode at a rate $\Gamma_L$ and leave the right quantum dot via the right electrode at a rate $\Gamma_R$, as well as electronic dephasing processes due to phonons at a rate $\gamma$. The dissipators are of the standard Lindblad form $\mathcal{D}[\hat{\sigma}]\hat{\rho}=\hat{\sigma}^\dag\hat{\rho}\hat{\sigma}-\frac{1}{2}\{\hat{\sigma}\hat{\sigma}^\dag,\hat{\rho}\}$, and we have defined the creation operators $\hat{\sigma}_{L}^\dag=|L\rangle\!\langle 0|$ and $\hat{\sigma}_{R}^\dag=|R\rangle\!\langle 0|$. Finally, cavity losses at rate $\kappa$ are described by a coupling of the cavity to a heat bath of the form
\begin{equation}
\mathcal{L}_{\mathrm{cav}}\hat{\rho}=\kappa \bar{n} \mathcal{D}[\hat{a}]\hat{\rho}+\kappa (\bar{n}+1) \mathcal{D}[\hat{a}^\dag]\hat{\rho},
\end{equation}
where $\bar{n}$ is the mean equilibrium occupation of the cavity.

\begin{figure*}
\includegraphics[width=\textwidth]{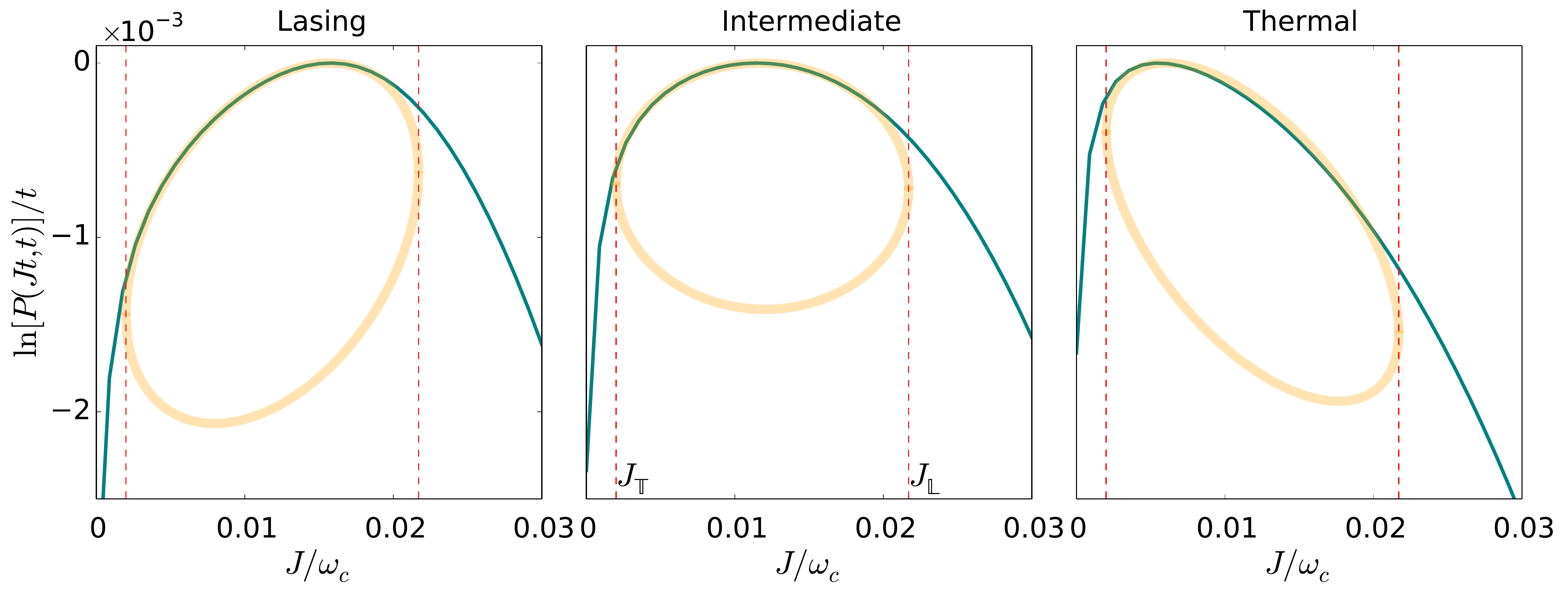}
\caption{(color online). \revision{Photon emission statistics}. The solid lines show exact results for the distribution $\ln[P(J t,t)]/t$ of the photon current $J$ in the lasing state, the intermediate regime, and the thermal state corresponding to the values of $\Gamma_R$ indicated in Fig.~\ref{fig:2}. Together with the exact results we show ellipses whose forms are determined by the switching rates in Fig.~\ref{fig:2}b. The ellipses are  delimited by the average emission rates in the lasing $J_\mathbb{L}$ and thermal $J_\mathbb{T}$ states (dashed lines). \revision{The fact that the distributions have the shape of a tilted ellipse demonstrates that the single-atom laser is bistable.}}\label{fig:3}
\end{figure*}

\emph{Wigner function.---} The state of the cavity is characterized by its density matrix $\hat{\rho}_{\mathrm{cav}}^{S}=\mathrm{Tr}_{\mathrm{elec}}\{\hat{\rho}^{S}\}$ with the stationary state determined from $\mathcal{L}\hat{\rho}^{S}=0$ and we trace over the electronic degrees of freedom. The cavity photons can be conveniently visualized using the Wigner distribution function shown in Fig.~\ref{fig:1}c for different values of the tunneling rates. We focus on the regime $\Gamma_R\ll\Gamma_L$, where electrons enter the DQD much faster than they leave it. With $\Gamma_R=0$, no photons are pumped into the cavity. As $\Gamma_R$ is increased, the cavity reaches the lasing state, where electronic transport takes place by exciting the cavity. With $\Gamma_R\ll\omega_c$, the broadening of the right level is smaller than the level splitting and the pumping mechanism is efficient. Due to the stochastic tunneling of electrons, the lasing state does not correspond to a coherent state with a fixed phase, but it is rather characterized by a ring-like Wigner function~\cite{supmat}. As $\Gamma_R$ is increased further, the right electronic level is broadened so much that electronic transport can take place without exciting the cavity. We then observe a crossover via an intermediate regime to a thermal state governed by an effective temperature determined by the electronic transport in the DQD~\cite{Ashhab2009}. Considering only the Wigner functions, the nature of the intermediate regime is not clear, but we note that similar crossovers have been observed in nano-electromechanical systems~\cite{Novotny2003,Novotny2004,Rodrigues2007,Harvey2008}.

\emph{Current \& noise.---} The Wigner function is a static observable which does not capture dynamical properties of the system. To better understand the dynamics, we consider time-integrated quantities such as the current and the noise in the DQD. These can be written as~\cite{Flindt2004}
\begin{gather}
\langle\hat{I}\rangle = e\langle\!\langle\tilde{0}|\mathcal{I}|0\rangle\!\rangle,\nonumber\\
S = e^2\left(\langle\!\langle\tilde{0}|\mathcal{I}|0\rangle\!\rangle-2\langle\!\langle\tilde{0}|\mathcal{I}\mathcal{R}\mathcal{I}|0\rangle\!\rangle\right),\nonumber
\end{gather}
where $\mathcal{I}$ is the super operator for the particle current running into the right lead, defined by its action on a density matrix as $\mathcal{I}\hat{\rho}=\Gamma_R\hat{\sigma}_{R}\hat{\rho}\hat{\sigma}_{R}^\dag$~\cite{Flindt2004,Flindt2005}. In addition, we use the notation $\langle\!\langle\tilde{0}|\mathcal{I}|0\rangle\!\rangle=\mathrm{Tr}\{\mathcal{I}\hat{\rho}^{S}\}$ and $\mathcal{R}$ is the pseudo-inverse of $\mathcal{L}$~\cite{Flindt2004,Flindt2005,Flindt2008,Flindt2010}. With these expressions at hand, we calculate the current and noise in Fig.~\ref{fig:2}a. The current is zero with $\Gamma_R=0$ \revision{(not shown)}. As $\Gamma_R$ is increased, we reach the lasing state with a large electrical current. The average current falls off as we go through the crossover to the thermal state by further increasing $\Gamma_R$, before it eventually converges  to the value of the current without the cavity~\cite{Gurvitz1996}. The current falls off with increasing $\Gamma_R$ due to dephasing and charge localization on the quantum dots because of the coupling to the right electrode~\cite{Gurvitz1996,Childress2004,Ashhab2009}. The crossover is also reflected in the Fano factor $F=S/e\langle I\rangle$ which displays a large peak in the intermediate regime. This we take as a \revision{possible} signature of a bistability in the \revision{single-atom laser}.

\emph{Bistability.---} \revision{To investigate if the single-atom laser is in fact bistable} we consider a model where \revision{the laser} switches slowly between the lasing state $(\mathbb{L})$ and the thermal state~$(\mathbb{T})$ with the unknown switching rates $\Gamma_{\mathbb{T}\leftarrow \mathbb{L}}$ and $\Gamma_{\mathbb{L}\leftarrow \mathbb{T}}$. This can be expressed by the master equation
\begin{equation}
\dot{\mathbf{p}}(\boldsymbol{\chi})=\mathbf{M}(\boldsymbol{\chi})\mathbf{p}(\boldsymbol{\chi}),
\label{eq:effmod1}
\end{equation}
where the rate matrix \revision{reads}
\begin{equation}
\mathbf{M}(\boldsymbol{\chi})=\left[
             \begin{array}{cc}
               \mathcal{H}_\mathbb{L}(\boldsymbol{\chi})-\Gamma_{\mathbb{T}\leftarrow \mathbb{L}} & \Gamma_{\mathbb{L}\leftarrow \mathbb{T}} \\
               \Gamma_{\mathbb{T}\leftarrow \mathbb{L}} & \mathcal{H}_\mathbb{T}(\boldsymbol{\chi})-\Gamma_{\mathbb{L}\leftarrow \mathbb{T}} \\
             \end{array}
           \right]
\label{eq:effmod2}
\end{equation}
\revision{and the vector $\mathbf{p}(\boldsymbol{\chi})=[p_\mathbb{L}(\boldsymbol{\chi}),p_\mathbb{T}(\boldsymbol{\chi})]^T$ contains the probabilities of being in the state $\alpha=\mathbb{L},\mathbb{T}$. We have also included the vector of counting fields $\boldsymbol{\chi}=[\chi_e,\chi_p]^T$ that couple to the number of transferred electrons and the number of emitted photons. The (unknown) generators $\mathcal{H}_\mathbb{\alpha}(\boldsymbol{\chi})$ describe the fluctuations in the uncoupled states.}

\revision{Assuming that this master equation captures the correct physics}, we can calculate analytically the mean current $\langle\hat{I}\rangle$ and the shot noise $S$ based on Eqs.~(\ref{eq:effmod1},\ref{eq:effmod2}) and from those expressions isolate the switching rates
\begin{equation}
\Gamma_{\alpha\leftarrow \beta}=\frac{2e}{S}\frac{\left(\langle\hat{I}\rangle-I_\beta\right)\left(\langle\hat{I}\rangle-I_\alpha\right)^2}{I_\alpha-I_\beta},
\label{eq:switchingrates}
\end{equation}
\revision{where $I_\alpha$ is the average electrical current in state $\alpha=\mathbb{L},\mathbb{T}$.} We have assumed that the low-frequency noise is mainly due to the switching process and have thus disregarded the fluctuations in the individual states. Using the exact results for the current and the noise in Fig.~\ref{fig:2}a, we then extract the switching rates shown in Fig.~\ref{fig:2}b.

\revision{\emph{Photon emission statistics.---} To confirm our picture of a bistability we now evaluate the photon emission statistics. At long times, the distribution of the photon current $J$ can be found from Eqs.~(\ref{eq:effmod1},\ref{eq:effmod2}) in a saddle-point approximation. This procedure yields} \cite{Jordan2004}
\begin{equation}
\frac{\ln P(J t,t)}{t}\simeq \frac{\left(\sqrt{\Gamma_{\mathbb{L}\leftarrow \mathbb{T}}|J-J_\mathbb{L}|}\pm\sqrt{\Gamma_{\mathbb{T}\leftarrow \mathbb{L}}|J-J_\mathbb{T}|}\right)^2}{J_\mathbb{T}-J_\mathbb{L}},
\nonumber
\end{equation}
\revision{where $J_\alpha$ is the average photonic current in state $\alpha=\mathbb{L},\mathbb{T}$. The saddle-point approximation describes the upper part of a tilted ellipse, corresponding to the minus sign above, while the lower part is given by the plus sign. The upper part of the ellipse is expected to capture} the photon emission statistics \revision{in the range} $J_\mathbb{T}<J<J_\mathbb{L}$. The tails of the distribution should be determined by the fluctuations in the individual states governed by the generators $\mathcal{H}_\alpha(\boldsymbol{\chi})$. The form of the ellipse is determined by the switching rates $\Gamma_{\alpha\leftarrow \beta}$, which can be controlled by modulating the electronic transport in the DQD.

In Fig.~\ref{fig:3} we show \revision{the photon emission statistics} based on the full GME in Eq.~(\ref{eq:GME}). Together with our numerically exact results obtained following Ref.~\cite{Flindt2010}, we show the tilted ellipse with switching rates taken from Fig.~\ref{fig:2}b. Figure~\ref{fig:3} confirms our picture of a bistability. In the range $J_\mathbb{T}<J<J_\mathbb{L}$, \revision{the distributions are well-described by the ellipses, demonstrating that the single-atom laser is bistable.} The tilt of the ellipses is given by the ratio of the rates $\Gamma_{\mathbb{L}\leftarrow \mathbb{T}}/\Gamma_{\mathbb{T}\leftarrow \mathbb{L}}$, while the width is governed by the sum $\Gamma_{\mathbb{L}\leftarrow \mathbb{T}}+\Gamma_{\mathbb{T}\leftarrow \mathbb{L}}$. The tails of the distribution are dominated by the statistical properties of the individual states. In combination, these building blocks provide us with a detailed understanding of the photon emission statistics from the solid-state single-atom laser.

\begin{figure}
\includegraphics[width=0.8\columnwidth]{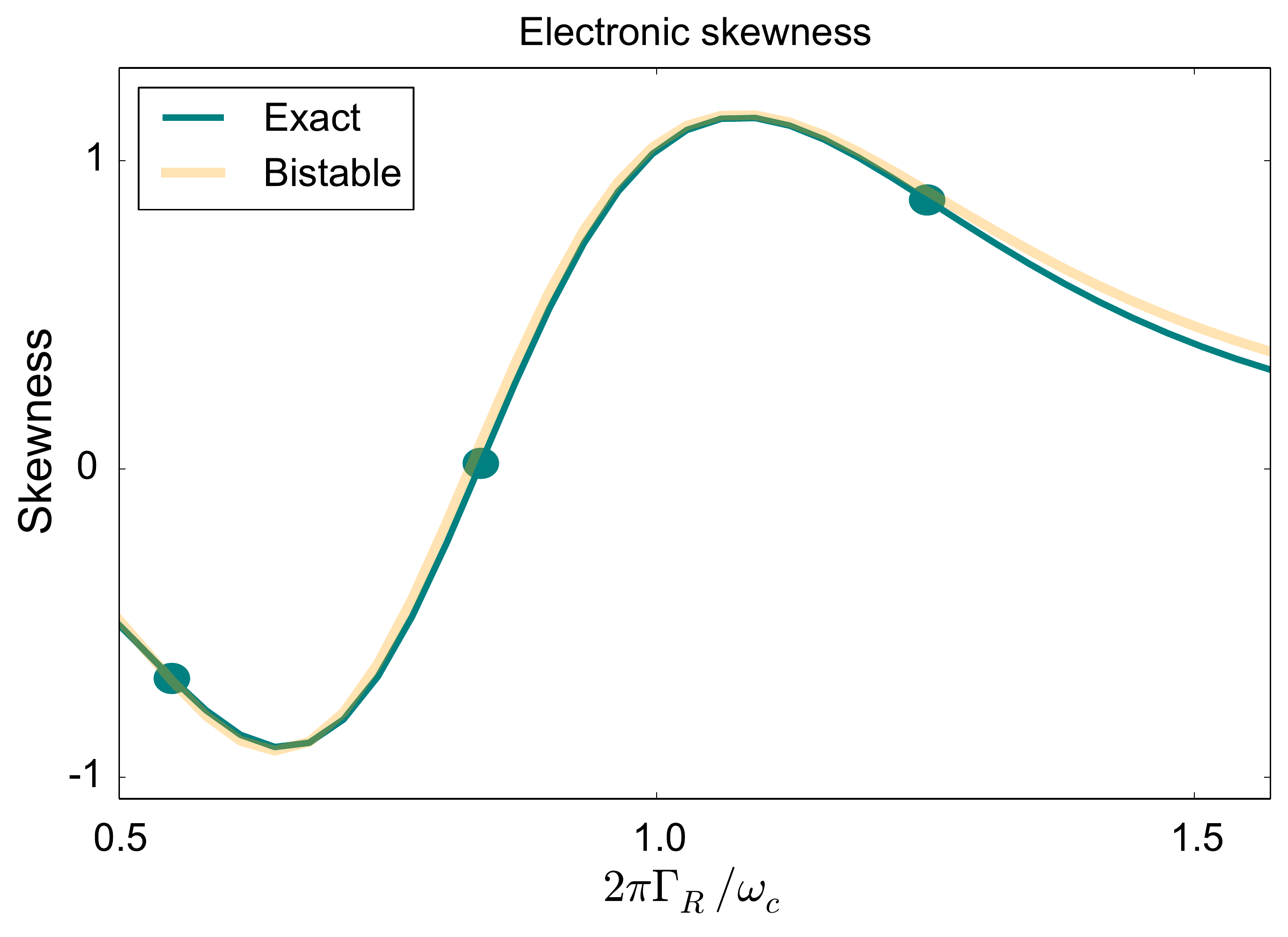}
\caption{(color online). \revision{Electronic skewness. The green line shows the skewness $\langle(\hat{I}-\langle\hat{I}\rangle)^3\rangle$. The yellow line is the skewness for a bistable system, $\langle(\hat{I}-\langle\hat{I}\rangle)^3\rangle=6(I_\mathbb{T}-I_\mathbb{L})^3\Gamma_{\mathbb{T}\leftarrow \mathbb{L}}\Gamma_{\mathbb{L}\leftarrow \mathbb{T}}(\Gamma_{\mathbb{L}\leftarrow \mathbb{T}}-\Gamma_{\mathbb{T}\leftarrow \mathbb{L}})/(\Gamma_{\mathbb{L}\leftarrow \mathbb{T}}+\Gamma_{\mathbb{T}\leftarrow \mathbb{L}})^5$~\cite{Flindt2005}. The agreement between the curves constitutes additional evidence for the bistability based on electrical measurements only.}}\label{fig:4}
\end{figure}

\emph{Experimental \revision{realization}}.--- \revision{For typical resonator frequencies, $\omega_c\simeq 5$ GHz, the parameters used here are close to those from Refs.~\cite{Delbecq2011,Delbecq2013,Frey2012,Frey2012b,Petersson2012,Toida2013,Wallraff2013,Viennot2014,Liu2014,Liu2015,Stockklauser2015,Viennot2015,Deng2014}.} \revision{The corresponding currents in Fig.~\ref{fig:2}a are then roughly 10~pA, which is well within experimental reach. Shot noise measurements are performed in the kHz regime to avoid flicker noise. This is sufficiently slow to capture the switching behavior governed by the MHz rates in Fig.~\ref{fig:2}b. To measure the photon emission distribution~\cite{Koshino2015,Koshino2015b}, statistics must be collected on time scales that are much longer than the inverse switching rates, in practice over several microseconds. If the emission statistics cannot be accurately sampled, the bistability can also be confirmed by measuring the electronic skewness shown in Fig.~\ref{fig:4} together with the skewness for a bistable system. The good agreement provides additional evidence of the bistability.}

\emph{Conclusions.---} \revision{We have predicted a bistability in a solid-state single-atom laser comprised of a DQD coupled to a waveguide resonator. The bistability can be demonstrated experimentally by measuring the non-trivial photon emission statistics which should take the shape of a tilted ellipse. Estimates of relevant time and frequency scales indicate that our prediction is within experimental reach. The single-atom laser does not need to be in the strong-coupling regime, and smaller couplings can be compensated for by increasing the cavity quality factor (and thus enhancing the lasing effect). Our analysis shows that the photon emission statistics can be modulated by electrical means by regulating the transport in the quantum dots. These findings are important for the controlled operation of solid-state single-atom lasers.}

\emph{Acknowledgments.---} We thank M.~Cirio, M.~Delbecq, C.~Emary, D.~S.~Golubev, I.~Khaymovich, V.~F.~Maisi, and P.~Zhang for useful discussions. Numerical calculations were performed with QuTip \cite{Johansson2012,Johansson2013}. FN is partially supported by the RIKEN iTHES Project, MURI Center
for Dynamic Magneto-Optics, the Impact program of JST, and a Grant-in-Aid for Scientific Research (A). CF is affiliated with Centre for Quantum Engineering at Aalto University.

\end{document}